# Computer Vision Methods for the Microstructural Analysis of Materials: The State-of-the-art and Future Perspectives

[1]Khaled Alrfou, [*2]Amir Kordijazi, [1]Tian Zhao

[1] Department of Electrical Engineering and Computer Science, University of Wisconsin Milwaukee, WI 53211 USA

[2] Colleges of Nanoscale Science and Engineering, SUNY Polytechnic Institute, Albany, NY 12203, USA

## Abstract

Finding quantitative descriptors representing the microstructural features of a given material is an ongoing research area in the paradigm of Materials-by-Design. Historically, the microstructural analysis mostly relies on qualitative descriptions. However, to build a robust and accurate process-structure-properties relationship, which is required for designing new advanced high-performance materials, the extraction of quantitative and meaningful statistical data from microstructural analysis is a critical step. In recent years, computer vision (CV) methods, especially those which are centered around convolutional neural network (CNN) algorithms have shown promising results for this purpose. This review paper focuses on the state-of-the-art CNN-based techniques that have been applied to various multi-scale microstructural image analysis tasks, including classification, object detection, segmentation, feature extraction, and reconstruction. Additionally, we identified the main challenges with regards to the application of these methods to materials science research. Finally, we discussed some possible future directions of research in this area. In particular, we emphasized the application of transformer-based models and their capabilities to improve the microstructural analysis of materials.

[*]**Corresponding Author**: Amir Kordijazi, Colleges of Nanoscale Science and Engineering, SUNY Polytechnic Institute, Albany, NY 12203, USA, email: kordija@sunypoly.edu, phone number: +1 414-629-0591

**List of Acronyms**

| | |
|---|---|
| 2DMOINet | 2D Material Optical Identification Neural Network |
| AFM | Atomic Force Microscope |
| AHSS | Advanced High Strength Steel |
| CAM | Class Activation Mapping |
| CCDCGAN | Constrained Crystals Deep Convolutional GAN |
| CFs | Carbon Fibers |
| cGAN | Conditional Generative Adversarial Network |
| CNN | Convolutional Neural Network |
| CT | Computer Tomography |
| CV | Computer Vision |
| CycleGAN | Cycle Generative Adversarial Network |
| DCNN | Deep CNN |
| DenseNet | Densely Connected Convolutional Network |
| DFT | Density Functional Theory |
| DL | Deep Learning |
| DLSP | Deep Learning for Structure Property |
| DSC | Dice Similarity Coefficient |
| EBSD | Electron Back-Scatter Diffraction |
| FCN | Fully Convolutional Network |
| GAM | Grain Average Misorientation |
| GAN | Generative Adversarial Network |
| GAP | Global Average Pooling |
| GIMP | GNU Image Manipulation Program |
| GMM | Gaussian Mixture Model |
| Grad-CAM | Gradient-Weighted CAM |
| GRF | Gaussian Random Field |
| GT-DNN | Graphene Trained Deep Neural Network |
| HRTEM | High-Resolution Transmission Electron Microscopy |
| I-GOS | Integrated Gradients Optimization Saliency |
| IoU | Intersection Over Union |
| kNN | k-Nearest Neighbors |
| MDD-Net | Material Defect Detection Network |
| ML | Machine Learning |
| MLP | Multi-Layer Perceptron |
| MVFCNN | Max-Voted FCNN |
| NASNet | Neural Architecture Search Network |
| NGO | Non-Grain-Oriented |
| OM | Optical Microscope |
| R-CCN | Region-based CNN |
| RF | Random Forest |
| RolAlgin | Region of Interest Align |
| SEM | Scanning Electron Microscope |
| SLIC | Simple Linear Iterative Clustering algorithm |
| SRGAN | Super-resolution Generative Adversarial Network |
| SRResNet | Super-Resolution Residual Network |
| SVM | Support Vector Machine |
| TEM | Transmission Electron Microscopy |
| TNT | Transformer in Transformer |
| UHCS | UltraHigh Carbon Steel |
| VAE | Variational AutoEncoder |
| ViT | Vision Transformer |
| VLAD | Vector of Locally Aggregated Descriptors |
| WDD-NET | Wafer Defect Detection Network |
| WGAN | Wasserstein Generative Adversarial Network |

# 1. Introduction

In the paradigm of Materials by Design, materials are designed from the atomic to the macroscopic scales, for a specific set of properties needed for a given performance [1]. By establishing linkage between processing-structure-properties (PSP) of materials, this approach can be instrumental in the discovery of new advanced materials, urgently needed for application such as more efficient renewable energy and energy storage technologies, reusable and biodegradable materials, and critical infrastructure materials [2]. This area of research has advanced significantly in recent years thanks to the expanded use of computational tools, the development of materials databases, improvements in experimental techniques, and the integration of artificial intelligence and machine learning algorithm into materials discovery research. Despite its recent achievements, the materials by design paradigm is still relatively new; several issues need to be resolved for further growth and impact [3], [4].

One main issue in the materials by design research is finding representative microstructural features, needed for building accurate PSP models. Although measuring microstructural features were developed more than a century ago, the integration of this rich knowledge base in the materials by design research has been a hurdle. This is due to the fact that the microstructural analysis mostly relies on qualitative, rather than quantitative, descriptions [5]. However, for constructing accurate PSP models, the extraction of quantitative and meaningful statistical data from microstructural analysis is essential [6]. Although numerous techniques for microstructure characterization and reconstruction have been developed [7], they do not apply to the materials by design approach. This is on account of significant information loss during representing microstructure using these techniques [8].

Microstructure informatics is an area of research that focuses on the development of data science algorithms in order to mine digital representation of material's hierarchical internal structure [9], [10]. The microstructure of a material can be described in a variety of ways. The crystal structure, the grain size and shape, and the orientation of the grains are examples of common metrics. In addition, other metrics describe defect distributions such as porosity and dislocation substructures. As it is widely accepted, all these characteristics have the potential to impact the material properties [11]. Computer vision (CV) methods and more specifically deep learning techniques have demonstrated promising results for developing tools that quantitatively and objectively collect rich and comprehensive microstructural information, required for building robust and concise structure-properties models [5], [12], [13].

This paper focuses on the application of various computer vision techniques for the microstructural analysis of materials. First, we overview some of the state-of-the-art CV techniques that have found application in materials science research. Then, we introduce the main image analysis tasks that can be performed using these techniques for different microstructural analysis datasets. Finally, we discuss some challenges regarding the application of CV methods, and possible future research direction to further exploit the extensive capabilities of these advanced methods in the materials discovery research.

## 2. Background

## 2.1. Deep Learning and CNNs

Deep learning (DL) methods have been applied in numerous fields, such as medical, health, biology, and material science, for feature learning and pattern classification [14] [15]. Past research showed that deep learning can outperform traditional machine learning approaches in several fields like image processing and computer vision [14]. This section briefly overviews deep learning concepts and discusses the most popular CNN (Convolutional Neural Network) techniques in material sciences.

Convolutional neural network (CNN) architecture is a powerful family of neural networks designed to work with images. It is among the most popular deep learning techniques and ubiquitous in the field of CV [16]. This is because of its success in solving challenging image classification problems previously solved by hand-crafted features [17]. In recent years, various enhanced CNN models have outperformed the state-of-art methods in a wide variety of visual recognition tasks [17] [18] [19] [20].

There are three distinctive characteristics of a CNN: local receptive fields, shared weights, and pooling (or sub-sampling). These characteristics help to identify similar local features and reduce the number of model parameters. The idea is not to connect the network fully but to replicate neurons at many places within a layer. A CNN uses various filters (or kernels) to apply convolution operations to the image pixels and generate feature maps that are layered on top of each other (Figure 1). Each feature map identifies a particular pattern relevant to a layer. Pooling is used to reduce the dimensions of feature maps and the number of network parameters. In the pooling layer, only a part of the previous layer is passed to the next layer, as shown in Figure 2 (a). This lowers the computational load and increases the translational invariance by reducing the impact of the precise locations of visual features [21].

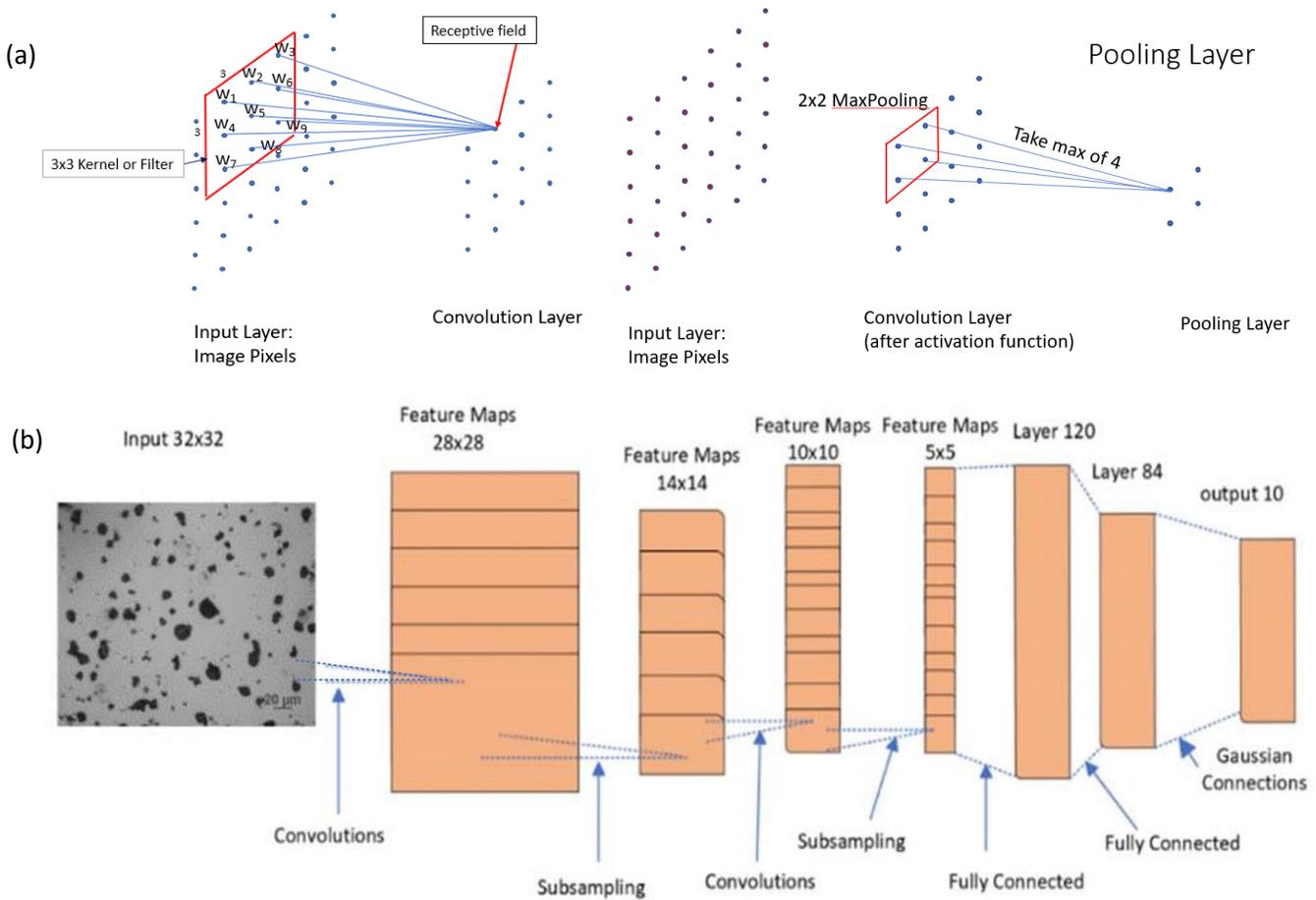

*Figure 1: (a) Two typical layers in a CNN: convolution layer and max-pooling layer. (b) the architecture of a simple CNN for image classification. The input image is a grayscale image with 32 x 32-pixel size fed to the CNN. The first convolution layers applied different kernels (filters) to produce 28 x 28 feature maps. The pooling layer (subsampling) used a window of size 2 x 2 and take average or maximum value in that window to reduce the dimensionality of the feature maps from 28 x28 to 14 x14. This block is followed by another convolution layers to produce 10 x 10 feature maps, and 2 x 2 pooling layer to reduce the dimensionality of the feature maps to 5 x 5. Then the feature maps are flattened to get a 1-D vector which was fed to two fully connected layers with 120 and 84 neurons, respectively, finally producing the output value.*

In the last decade, numerous CNN architectures have been developed for image analysis tasks such as image classification [17] [22] [25], image segmentation [26], object detection, and instance segmentation. For example, AlexNet [17] used five convolution layers for image classification tasks. VGG16 [22] increased the network depth to 16 weight layers using very small convolution layers of 3x3 filters. ResNet [20] was designed to address the vanishing gradient problem and proposed the residual connection. Following that, several variations of ResNet have been proposed [23]-[24]. These architectures achieved high prediction accuracy on datasets such as ImageNet. For image segmentation, different types of CNN have been developed. Fully convolutional network (FCN) is an architecture mainly for semantic segmentation and can be efficiently trained to make dense predictions for per-pixel tasks [26]. The main idea of FCN is to replace the fully connected layers in a CNN with convolutional layers to enable a classification net that outputs a spatial map for pixel-wise learning. FCN consists of down-sampling (convolution), up-sampling (deconvolution), and skip connection layers. The skip connection layer is used to combine the coarse, semantic, and local appearance information to get accurate and fine predictions [26] [27]. To improve accuracy, variations of FCNs were introduced. For example, U-Net is a symmetric encoder and decoder network with skip connections. The U-Net architecture has many channels and skip connections to prevent information loss when reducing image resolution. SegNet [28] is a convolutional encoder-decoder network for image segmentation as well. The encoder network in the SegNet has the same topology as the convolutional layers of VGG16 without fully connected layers. The decoder part used the memorized max-pooling indices received from the corresponding encoder to perform non-linear up-sampling of their input feature maps. This step is used to produce sparse feature maps, which are convolved with a trainable filter bank to produce dense feature maps. The decoder outputs feature maps fed to a SoftMax classifier for pixel-wise classification [28]. For object detection, Region-based CNN (R-CCN) is used to detect the location of a set of objects. It has three steps: (1) generating region proposals using the selective search algorithm, (2) extracting a feature vector from each region proposal, and (3) using a pre-trained Support Vector Machine (SVM) algorithm to classify the region proposal to either the background or one of the object classes [29].

## 2.2. Generative Models:

The generative machine learning methods were used in applications such as image super-resolution and microstructure reconstruction. The generative adversarial network (GAN) and variational autoencoder (VAE) are two particularly powerful methods.

1- **Generative adversarial networks (GANs)**

GAN contains a generator and a discriminator, which are trained against each other in an unsupervised learning process. As shown in Figure 2, the generator generates fake data by transforming the latent variables $z$ with a prior distribution $p_z(z)$ to samples close to actual data. The discriminator learns to distinguish the real data from the fake ones produced by the generator. After the training completes, the discriminator should be unable to determine fake data from real data. The training is performed using mini-max objective learning as a loss function (or adversarial loss function) [30] [31]. The adversarial loss function is flexible and can be modified for other purposes. For example, Wasserstein GAN (WGAN) is designed to improve learning stability [32]. Super-resolution GAN (SRGAN) is developed for creating super-resolution images [4] [33]. CycleGAN is trained for image-to-image translation models [34]. Conditional GAN (cGAN) sets the conditional variables for both the generator and discriminator during the training [35].

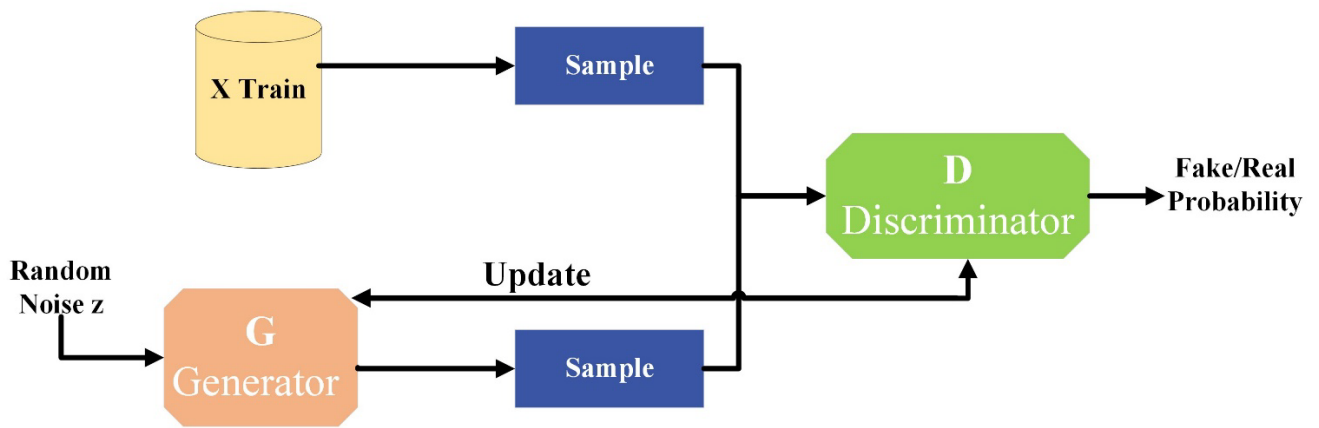

*Figure 2: GAN architecture includes a generator G and a discriminator D, where the generator learns to output sample (fake) data similar to the training data. At the same time, the discriminator is trained to distinguish the generated data from the training (real) data [30].*

2- **Variational autoencoder (VAE)**

Variational autoencoder (VAE) gives control over the data generated using latent variables. As shown in Figure 3, an autoencoder (AE) model includes an encoding and a decoding network. In VAE, the encoder part maps its input into a probability distribution, where the encoder learns the latent variables from the input images and the decoder takes the samples' latent variables to generate new images. VAE assumes that the data is generated from a multivariate normal distribution $N \sim (\mu, \sigma)$, where the mean and standard deviation are two vectors [36], [37].

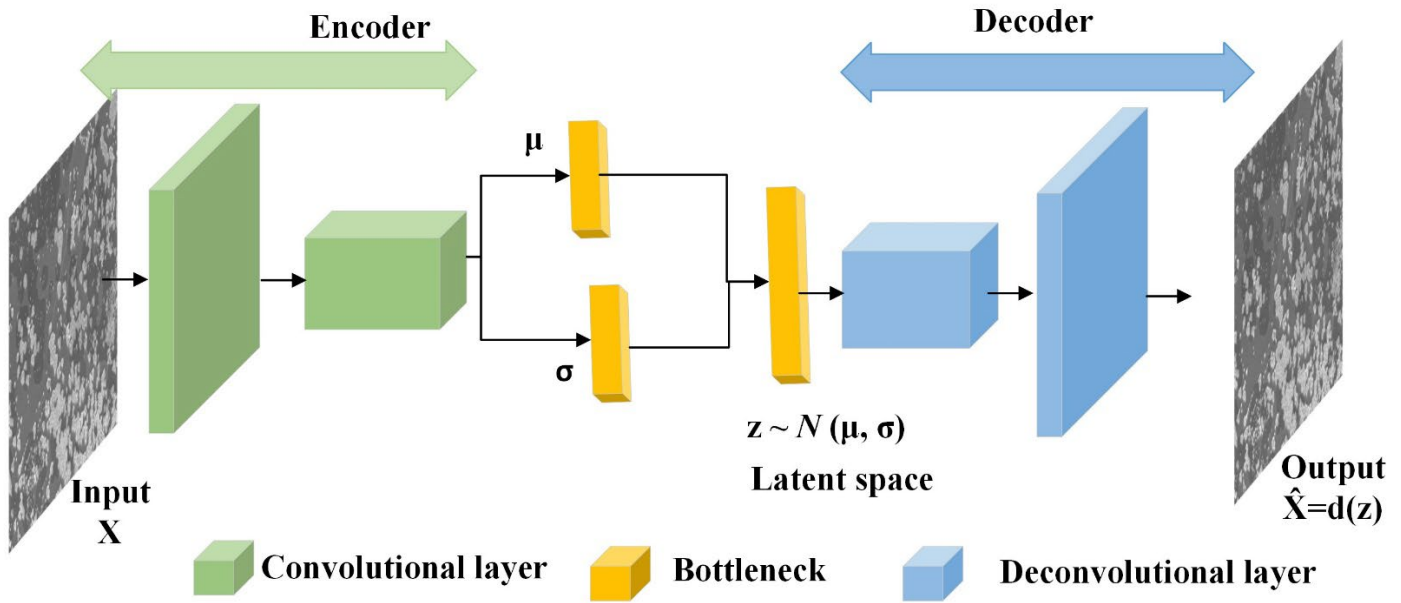

*Figure 3: The variational autoencoder (VAE) architecture. The input image map to a probability distribution over the latent space. Each point in the latent space is sampled from the probability distribution. The decoder takes the sample latent variable to generate a new image [24].*

## 2.3. Transformers in CV

The core idea of CNN is its convolution layer, which operates locally and provides translational equivariance. However, the local receptive field in a convolution layer is limited in capturing the long-range relationship between pixels. The transformer-based models are solutions to this problem by capturing the long-range pixel relationships and the long-range dependencies within the input. The transformer-based models have attracted great interests in CV domains [38] such as image recognition, image segmentation [39], object detection [40], [41], image super-resolution, and image generation [42]. Below, we briefly discuss the basic components of the transformer-based models.

The critical component of a transformer-based model is self-attention. The idea behind the self-attention mechanism is to learn self-alignment that provides the ability to model long-range dependencies between image patches. The multi-head attention consists of multiple self-attention layers, where each self-attention is called head attention [30]. A transformer-based model's training mechanism generally operates in two stages [38]. First, the pre-training stage is performed on a large-scale dataset in a supervised or a self-supervised way. Second, the fine-tuning stage adapts the pre-trained weights to the downstream tasks (e.g., image classification, image segmentation, object detection) by using small or mid-scale datasets.

Dosovitskiy *et al.*[38] introduced a Vision Transformer (ViT) that is built on a vanilla transformer network for Natural Language Processing (NLP) [43] with the fewest possible modifications to capture the global context of an input image. As shown in Figure 4, the ViT model architecture applied a pure transformer directly to sequences of image patches, flattened these patches, fed them to the linear projections, and provided the sequence of linear embeddings of these patches as an input to the transformer encoder model. The Multi-Layer Perceptron (MLP) head on the top of the model is used for the classification task. ViT was used for image classification tasks. The pre-training was performed using a large-scale dataset (14M-300M images) and was fine-tuned downstream for specific classification tasks using ImageNet dataset. Despite ViT's ability to capture the long-range dependencies between patches, it still suffers from poor performance in extracting local information. Additionally, the time complexity of the SoftMax function for each self-attention block is quadratic with respect to the length of the input sequence, which limits its applicability to high-resolution images. Recent research have focused on improving the transformer models to capture local information and resulted in new variations of the transformer-based models such as Transformer in Transformer (TNT) [44], Twins [45], Swin-Transformer [46], and Shuffle Transformer [47]. Table 1 summarizes the popular models discussed in this section and their advantages and limitations.

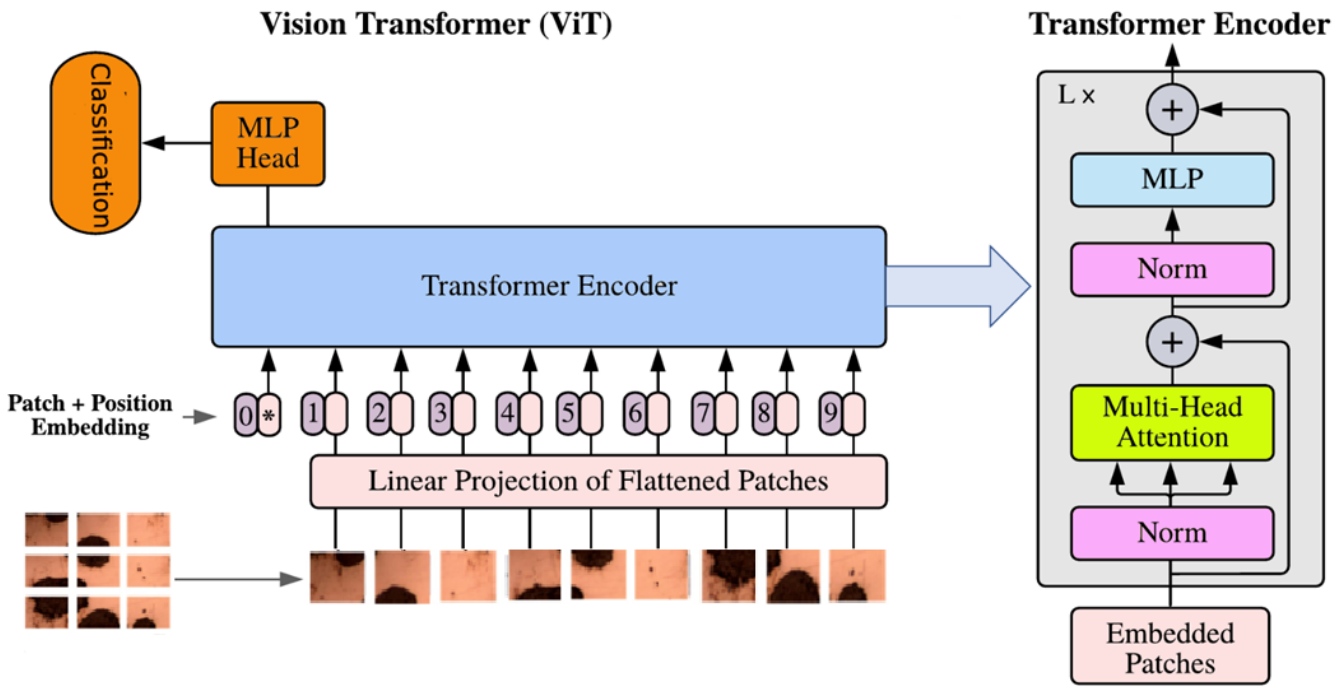

*Figure 4: Vision Transformer (ViT) model architecture. ViT breaks the image into patches and treats each patch as a token. The patches are flattened and fed to the linear projection. The linear projection produces the sequence of linear embeddings that are fed to the Transformer encoder for the training. The MLP is used for the classification task. [38]*

# 3. CV for Microstructural Analysis:

Digital image processing technology for microstructural images is widely applied in materials science fields. Depending on the resolution of interest, images might come from different image characterization techniques (for both micro and macro images). Image characterization techniques such as Atomic Force Microscope (AFM), Scanning Electron Microscope (SEM), Transmission Electron Microscope (TEM), Optical Microscope (OM) are utilized for microscopic image acquisition, while computer tomography (CT) is used for macroscopic image acquisition. These techniques allow us to understand the unique characteristics of the materials at various scales. Examples of phenomenon captured by these techniques are defect evolution, second-phase identification, dislocation migration, phase transformation, and analysis of three-dimensional structure. Recently, CV offered different techniques for extracting information from microstructural images, which can lead to the discovery of new materials with enhanced properties [48]–[50]. CV analysis tasks such as image classification, semantic segmentation, object detection, and instance segmentation have been used for defect detection, morphological feature analysis, chemical composition, and material design. In this section, we explore the recent published research about the application of CV in material science.

## 3.1. Image classification

Classification models assign a label to each image in a dataset. The label is the best representation for the image. For example, wafer-defect detection is one of the key challenges facing the semiconductor manufacturing companies. Chen et al. [51] proposed a novel CNN method called wafer defect detection network (WDD-Net) to detect wafer structural defects. The idea of WDD-Net is to reduce the number of parameters and calculations while preserving the depth of the network. As shown in Figure 5, the WDD-NET architecture consists of a 3×3 standard convolution and three depth-wise separable convolution (3×3 separation convolution and 1×1 standard convolution). The last layer was used for average pooling to ensure information integrity by taking global average pooling (GAP) while removing the fully connected layer. Two CNN architectures, VGG-16 and MobileNet-v2, were also trained for comparison purposes. The dataset has more than 10,000 grain images. The dataset was not balanced, where the qualified grains (positive data) were over 95% of a wafer and the defect grains (negative data) were only 5%. To enhance the generalization ability of the model, data augmentation was applied to increase the number of defect images using affine transformation and GAN. The results show that the three methods are able to detect wafer defects with accuracy of more than 99%. They concluded that VGG-16 and MobileNetv2 [52] were better than WDD-Net for the limited data sets, but WDD-Net reduces the number of parameters, computation, and the model size.

Mulewicz et al.[53] developed Deep CNN (DCNN) by fine-tuning ResNet18 (that was trained on ImageNet) to classify different types of microstructures of low-alloy steels. DCNN was used to classify 8 microstructure classes of steel grades: C15 (bainite with carbide), C45 (ferrite with grain boundaries and small islands of perlite), C60 (martensite), C80 (perlite), V33(ferrite with grain boundaries), X70 (granular bainite), X70 (columnar bainite), and carbide free steel. The dataset has 33,283 photograph images of different types of microstructures. The proposed method achieved an accuracy of 99.8%.

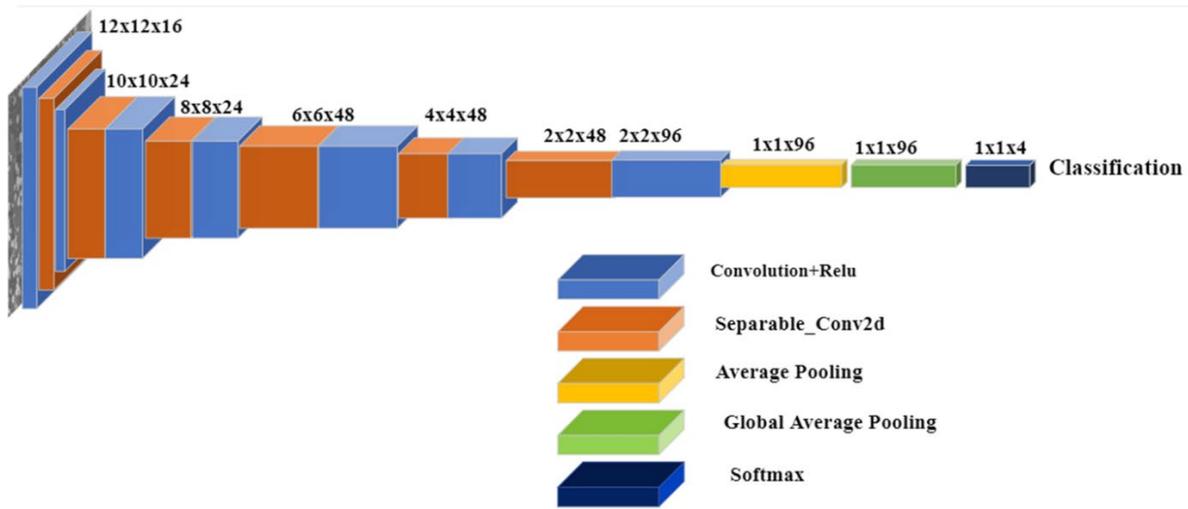

*Figure 5: WDD-Net architecture consists of standard convolution, depth-wise separable and global average pooling to reduce parameters and computation.[51]*

Elbana et al. [54] used six CNN models (VGG-16, VGG-19, ResNet50, Mobile-Net [55], Inception-V3, and Xception [56]) for the microstructural constituents of ultrahigh carbon steel (UHCS) to classify Spheroidite microstructural images. The open-source UHCS dataset was used in this research, consisting of 961 micrographs including Pearlite, Spheroidite, Cementite network, Pearlite containing Spheroidite, Widmanstatten cementite, and Martensite. There were 374 Spheroidite microstructural images in the dataset resulting from twenty-three distinct annealing schedules. The Spheroidite micrographs were divided into different classes based on different annealing conditions with the same magnification scale. Different datasets were constructed with distinct classes based on the magnification scales. Each image in each dataset was cropped to 224×224 sub-images. The six CNN models used transfer learning by either directly applying the pretrained model or fine-tuning the pretrained model. The results showed that CNNs incorporating pretrained models had the validation accuracies of 98.33% while the fine-tuned VGG16 model achieved the validation accuracy of 100%.

Banerjee and Sparks [57] implemented CNNs (VGG16, InceptionV3, and Xception), MLP, and Random Forest (RF) to classify microstructural images as dendritic or non-dendritic. The dataset consists of 133 dendritic and 444 non-dendritic structures from 21 alloy systems such as Al-Cu, Cu-Zn, Cast-Iron, and Carbon steel. They employed data augmentation via rotation and translation to increase the data points six-fold. The CNN methods used transfer learning by applying pre-trained networks of VGG16, InceptionV3, and Xception [56]. For MLP and RF, Binary Red Deer Algorithm (BRDA) optimization was employed for feature engineering and for improving the binary classification of microstructural images. The CNN methods with transfer learning achieved F1 scores of 80.1%–82.2% while MLP and RF achieved F1 scores in the range of 96%.

Baskaran et al [58] proposed a new paradigm for systematic segmentation of morphological features in titanium alloys. The proposed method used two-stage pipeline consisting of a classification step and a segmentation step. In the classification step, a CNN architecture consisting of three convolutional layers and one fully connected layer were developed to classify the microstructural images into three labels: lamellar, duplex, and acicular. The segmentation step used the watershed technique and the area fraction of the dominant-variant to process titanium microstructures. 1225 material microstructural images were used to train and test the CNN. The proposed method achieved an average accuracy of $93.00 \pm 1.17\%$.

### 3.2. Feature extraction

As we discussed in the background section, CNN architecture consists of convolutional layers to create and downsize feature maps and fully connected layers responsible for the classification of the image. CNN architecture can be employed as a feature extractor by removing the fully connected layers. In this format, each output vector represents a feature extracted from the microstructural image as shown in Figure 6. For example, after training VGG-16 using ImageNet and removing the last part of the model (fully connected layers), we can extract the features and feed them as inputs to the traditional machine learning models such as RF, Bayesian classifier, and Naive Bayes (NB). This framework is especially applicable to construct structure-properties relationships of materials.

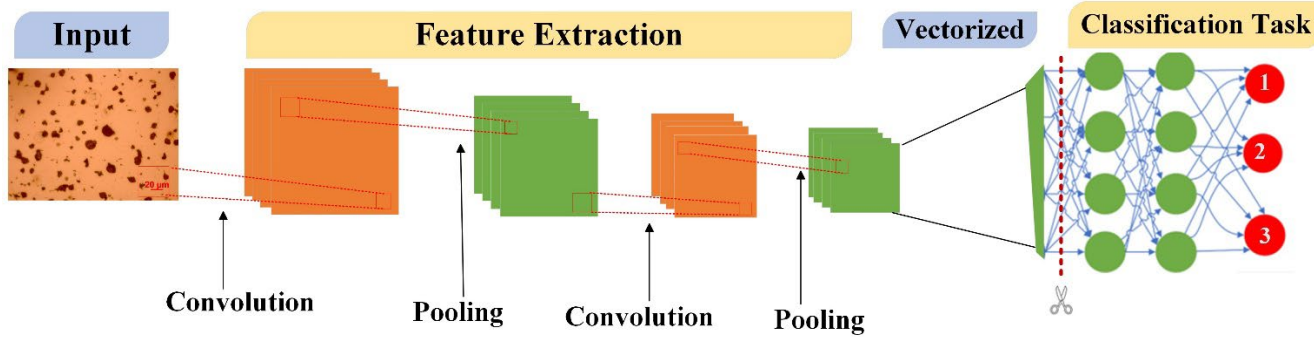

*Figure 6: Backbone of CNN for feature extractor. The feature maps can be extracted from CNN by removing the last part which is the fully connected layer. For example, these feature maps can be used to fed to a traditional ML for classification analysis task.*

Luo et al.[59] developed a CNN framework that automatically detects and classifies complex airborne carbon nanotubes/nanofibers (CNTs/CNFs) from TEM images. Airborne carbonaceous nanomaterials can have different physical–chemical properties and are very likely to form mixtures of individual nano-sized particles and micron-sized agglomerates with complex structures and irregular shapes, which makes the classification task extremely difficult. As shown in Figure 7, the framework was developed based on VGG16 that has five convolutional blocks with hypercolumn (Hcol)-based approach (VGG16+ 10% Hcol). Its result was compared with VGG16 with a linear SoftMax classifier (VGG16+ LS) that also has five convolutional blocks and fully connected layer. The architecture of the framework consists of the hypercolumns that were used to extract information from each pixel at different length scales. The length of the hypercolumn vectors is the sum of these length scales. The hypercolumn vectors concatenated the information from various VGG16 convolutional layers (2, 4, 6, 9, and 12). Each pixel represents 1472 features, and each image can cover over 73 million features compared to VGG16+ LS, which contains only 4096 features per image extracted from the fully connected layer. Hypercolumn feature vector was used to perform the classification of images, which includes three steps. The first step uses K-means to cluster the feature vectors in the high-dimensional space. The second step calculates the residuals for local features relative to the closest cluster centroids and concatenates the residuals to create Vector of Locally Aggregated Descriptors (VLAD) encoding. The third step tests three types of boosting algorithms (AdaBoost, random forest, and gradient boosting) for classification performance. The dataset has 5323 grey-scale TEM images of various nanostructured carbon samples. They defined five major classes representing the CNT/CNF structures: Cluster (Cl), Fiber (Fi), Matrix (Ma), Matrix-Surface (MS), and Non-CNT (NC). The sub-groups of NC classes are Graphene Sheets (GS), Soot Particles (SP), High-Density Particles (HDP), Polymer Residuals (PR), and Others (O). Since the data is not balanced, transfer learning and data augmentation have been applied to avoid overfitting. The proposed framework achieved 90.9% accuracy on the classification of a 4-class CNT dataset (including Cl, Fi, Ma, and MS) compared to VGG16+LS with 88% and 84.5% accuracy on a more complex 8-class dataset (including Cl, Fi, Ma, MS, GS, SP, HDP, and PR) compared to VGG16+LS with 80.2%. The framework can be also extended to the automated structural classification for other nanomaterials.

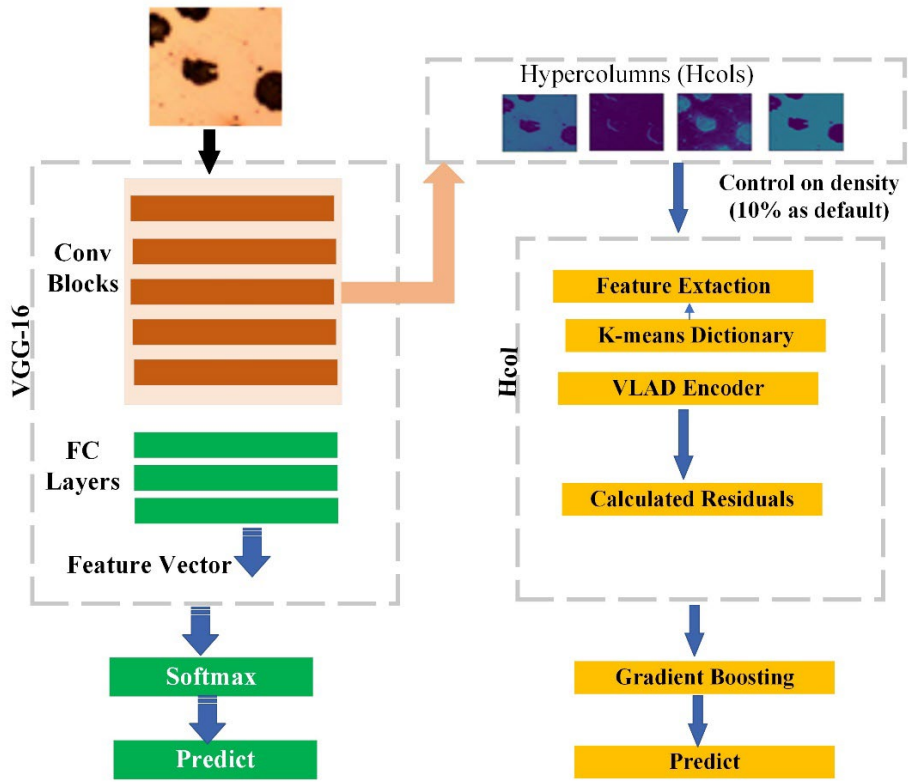

*Figure 7: The left is a VGG-16 image classification pipeline with a linear-SoftMax classifier (VGG-16+ LS). The right is the VGG-16+ (10% Hcol) architecture, where the hypercolumn extracted the feature information from VGG-16 followed by VLAD classifiers. [59]*

Ivo et al. [60] developed a method to classify photomicrographs of 1.26% Si non-grain-oriented (NGO) electrical steel. The electromagnetic efficiency of NGO electric steel is related to its grain size, which depends on the material's manufacturing process. However, it has been difficult to determine the efficiency of the electrical steel based on a particular manufacturing process. They proposed an intelligent, fast, and precise extractor-classifier combination method to classify NGO electrical steel using its photomicrographs by considering crystallographic texture and hysteresis curves data. Transfer learning using VGG, Neural Architecture Search Network (NASNet), Xception [56], MobileNet [55], ResNet, and Densely Connected Convolutional Network (DenseNet) [61] was applied as feature extractor and several traditional machine learning models (Bayesian classifier, Naive Bayes, MLP, kNN, Random Forest, and SVM) were employed as classifiers to classify data points using 127 images. The best performing extractor-classifier combination was the InceptionV3 architecture with the kNN method, which achieved 100% for both accuracy and F1-Score.

CV has been successfully employed for heterogeneous dataset. Xiang et al. [62] combined the multi-layer perceptron (MLP) and CNN to predict the creep rupture time of Fe–Cr–Ni heat-resistant from various centrifugally cast HP40Nb alloys (≈Fe– 25Cr– 35Ni- 1.5Nb- 0.4C wt%) tubes. The dataset consists of textual data (chemical composition and creep testing conditions) and visual (as-cast microstructure) images. MLP was trained to analyze the textual data and CNN was trained to analyze the visual data. The proposed method used fusion deep learning model by concatenating the textual and visual features extracted from MLP and CNN layers, respectively. It then fed the concatenated features to a densely connected multilayer to estimate the creep rupture time as its single output. The proposed method was trained in multi-source heterogeneous data, and it achieved a significant enhancement in prediction accuracy.

### 3.3. Semantic Segmentation

Semantic segmentation has been applied to various types of microstructural images. In semantic segmentation, a label is assigned to each pixel in an image, which provides information about the size, shape, and position of features of interest in an image. For example, Sadre et al.[63] compared two methods to perform the segmentation of complex features in phase-contrast high-resolution transmission electron microscopy (HRTEM) images of monolayer graphene. A conventional Bragg filtering algorithm and UNet architecture were used. UNet outperformed the Bragg filtering algorithm in all performance metrics including F1, Jaccard, precision, and recall scores. UNet was robust in preventing incorrect classification of structurally important regions compared with Bragg filtering. UNet achieved 97.9% accuracy compared with Bragg filtering with 95.6% accuracy.

Alrfou et al.[64] proposed a ML pipeline to create an automatic model to segment the graphite nodule in the microstructure of ductile iron. The ML pipeline consists of two steps. The first step was unsupervised learning using K-means clustering with manual thresholds to generate segmentation masks for the training dataset. This step significantly reduces the amount of time for preparing the training dataset. The step two was supervised learning by applying UNet to predict the percentage of graphite in the microstructure of ductile iron. To reduce memory usage and augment the training dataset, each 1280 × 964 input image is divided into 256 × 256 tiles. The algorithm achieved an Intersection Over Union (IoU) value of 96.3 %.

Decost et al. [65] applied PixelNet CNN with transfer learning (using a VGG16 network pretrained on ImageNet) for the Ultra High Carbon Steel Database. The database includes 24 micrograph images of the 645 × 484 resolution. They performed two segmentation tasks: semantic microconstituent segmentation of steel micrographs into four regions (grain boundary carbide, spheroidized particle matrix, particle-free grain boundary denuded zone, and Widmanstätten cementite) and segmenting cementite particles within the spheroidized particle matrix. The models were trained on 20 hand-annotated images using sixfold cross-validation with batch normalization, dropout, weight decay regularization, and data augmentation to avoid overfitting. Two loss functions were used: the standard cross-entropy classification loss and focal loss. For semantic microconstituent segmentation task, the predictions of model1, which used the focal loss function, achieved the IoU of 62.6 ± 2.5, while model2, which used cross-entropy loss function, achieved the IoU of 75.4 ± 3.7. For the spheroidite particle segmentation task, the model used focal loss function and achieved the IoU of 72.4 ± 3.1.

Azimi et al. [66] used pixel-wise segmentation of steel microstructure dataset. They applied an ensemble of fully convolutional neural network called Max-Voted FCNN (MVFCNN) to segment martensite, tempered martensite, bainite, and pearlite in carbon steels. As shown in Figure 8, the FCNN architecture is encoder-decoder architecture similar to VGG16 except that they replaced the fully connected layers with convolutional, up-sampling, and skip layers. The output of FCNN is a 3D matrix with the number of channels equal to the number of classes. A SoftMax layer is used to classify each pixel to the corresponding microstructure class $C_i$. In order to classify objects rather than pixel, max-voting policy was applied on each object, assigning it to the class of the majority of the pixels. The steel image dataset consists of 21 LOM and SEM images with 7000 × 8000 pixels. To reduce memory use, each image is cropped into 1000 × 1000 patches using sliding windows. Data augmentation was applied to the unbalanced dataset by rotating each patch by 90°, 180° and 270° to reduce overfitting. This method showed 93.94% accuracy.

Ajioka et al. [67] developed two DL methods using SegNet and UNet to perform segmentation for the microstructural images of ferrite-martensite dual phase steel. These methods were used to segment the ferrite phase, martensite phase, and ferrite grain boundary. The DL methods were compared to the existing methods such as segmentation by thresholds. The training and testing datasets consist of 40 and 10 Optical Microscopy (OM) images, respectively. It shows that DL methods are more accurate than the existing method. UNet achieved an accuracy of 95.10% for all phases compared to SegNet with 91.42% accuracy.

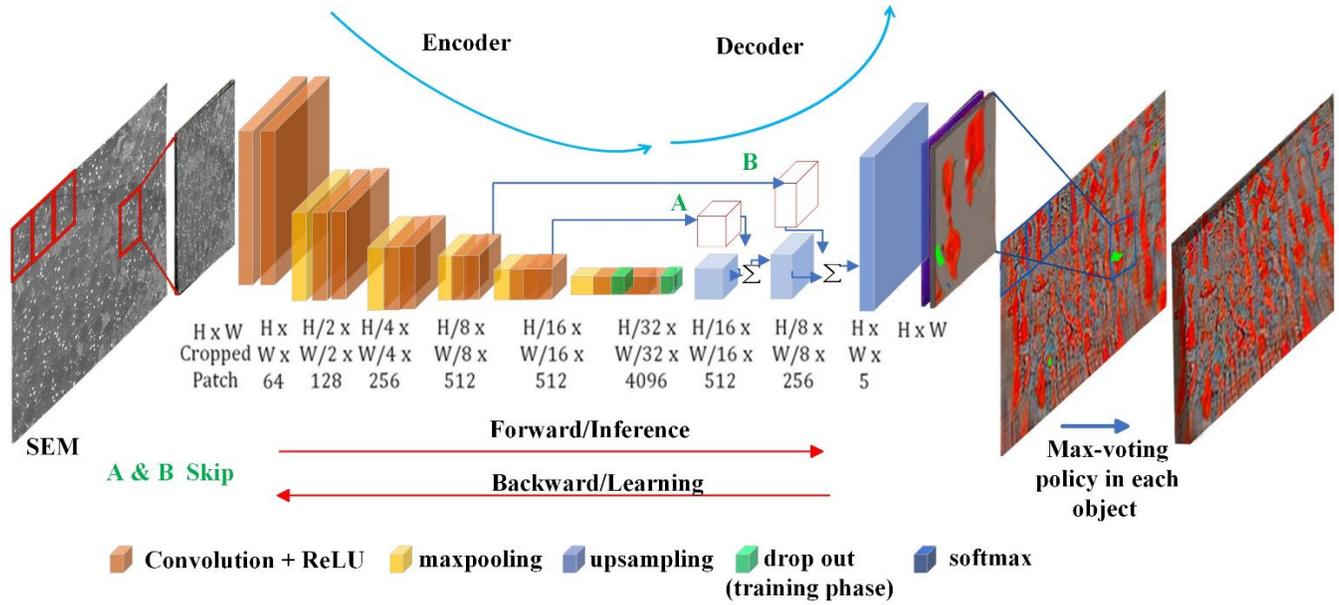

*Figure 8: Max-voted segmentation-based microstructural classification approach using FCNNs (MVFCNN) [66]*

Lu *et al.* [68] proposed a new segmentation model called *Swin-UNet++* to perform the automatic representation of dimples micro-features of an alloy fracture surface with complex morphology of 2.25Cr1Mo0.25V steel. The proposed method is based on Swin-UNet [69] and UNet++ [70]. Swin-UNet is a U-shaped architecture based on Swin transformer model. The Swin-Unet consists of encoders, a bottleneck, skip connections, and a decoder. Figure 9 shows the Swin-UNet++ architecture, where the decoder was redesigned to fuse additional feature representation by introducing a vice path to up-sample the feature representations from the last Swin-T block encoder. The proposed method can efficiently predict the boundaries and interior regions since it is based on Swin-UNet, which can capture the global and long-range dependency from the input image. Swin-UNet++ can accurately identify dimples, has a much higher prediction accuracy, and is more robust than Swin-uNet and UNet. It has 94.65% dice similarity coefficient (DSC) compared to Swin-UNet, which has 86.95% DSC.

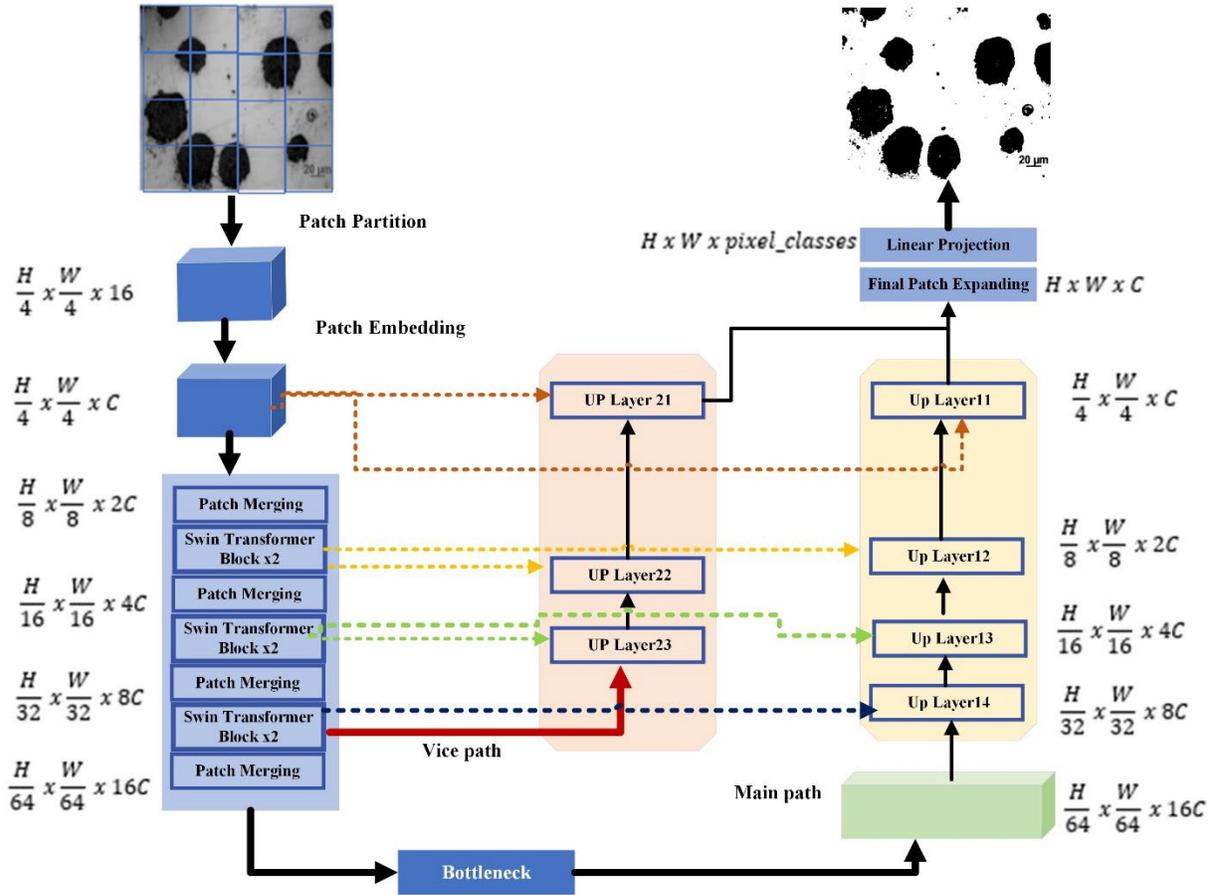

*Figure 9: Swin-UNet++ architecture consists of an encoder, bottleneck, skip connections, vice path and decoder. [68]*

## 3.4 Object detection and instance segmentation

Object detection is used to localize and recognize object instances in the image by defining the bounding boxes around the objects of interest. The output are object locations and the sizes of the bounding boxes. Localization is a specific case of detecting a single object in an image [71], while instance segmentation assigns a label to each instance of an object in an image. Object detection and instance segmentation have been applied to material microstructural images. For example, to predict the mechanical and physical properties of materials, Agbozo and Jin [72] applied Mask Region CNN (Mask R-CNN) to locate carbide particles, using bounding box indicators based on the concept of Region of Interest (ROI). The method used transfer learning (ResNet-101 trained on MS COCO dataset [73]). This method was also used for instance segmentation. It first finds the ROI by generating the bounding box of the blobs (Binary Large Objects), which is a group of connected pixels in the binary image that share common properties. It then calculates the areas and the perimeters of the blobs measured in pixels within each image. In this experiment, around 58 SEM images were used and the proposed method achieved over 90% accuracy.

Fu et al. [74] used Faster R-CNN to detect the dendrite cores of Ni-based superalloys with complex shapes of crystallographic symmetry. They applied data augmentation by rotating the image around the center of dendrite cores and applied transfer learning using VGG16. The experiments used 2D cross-sectioned microscopic images of Ni alloy samples. The method showed the effectiveness for both the detection accuracy and computational efficiency. It achieved 94.26% on mean average precision. Figure 10 shows a general Faster R-CNN architecture.

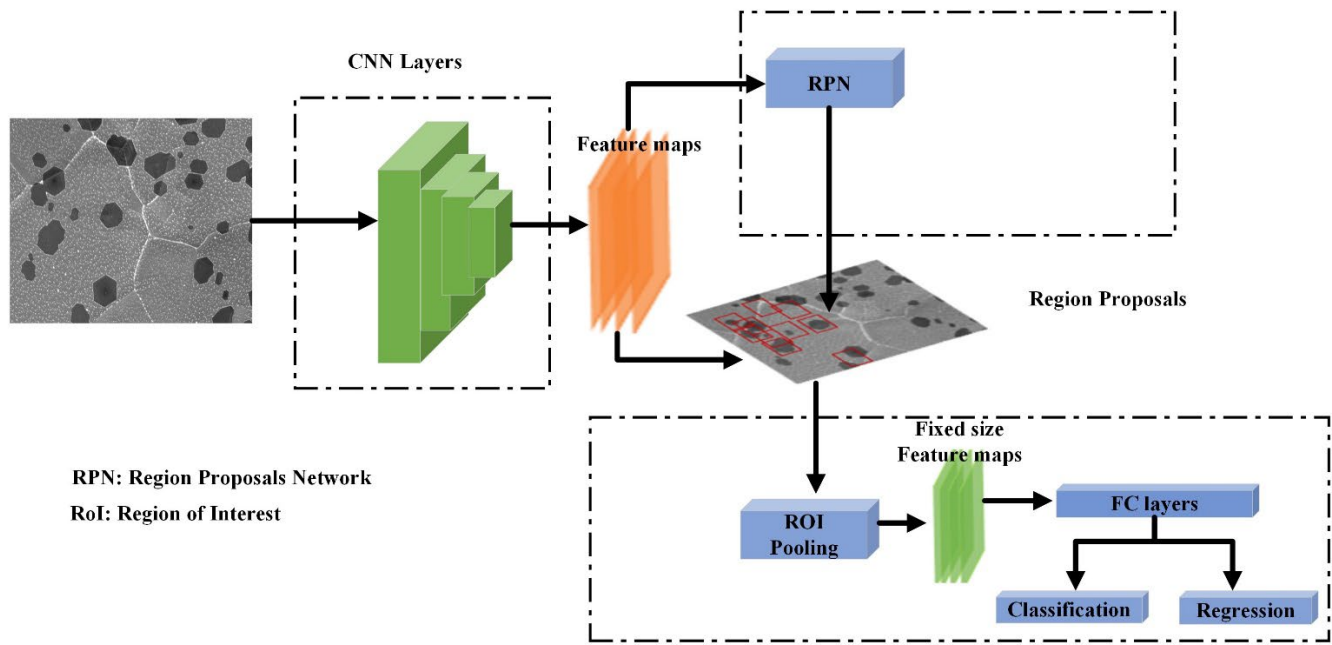

*Figure 10: General Faster R-CNN architecture. The image is fed into CNN only once to extract the feature map. RPN is a FCN used for generating region proposal. The proposal and feature map are fed to RoI pooling to extract fixed length feature map vector from each region proposal. The fixed feature map vector passed to FC layers to predict the class scores and the bounding box for each object [75].*

Masubuchi et al.[76] developed and implemented a deep learning-based segmentation in an autonomous robotic system for exfoliated 2D materials detection in a motorized optical microscope. Mask R-CNN was trained to detect various exfoliated 2D crystals on $SiO_2$/Si substrates (van der Waals heterostructures, graphene, hBN, $MoS_2$, and $WTe_2$). The detection method can process a 1024 × 1024 optical microscope image in 200 milliseconds, enabling the real-time detection of 2D materials and efficient detection of varying microscopy conditions such as illumination and color balance. Figure 11 shows the general Mask R-CNN architecture.

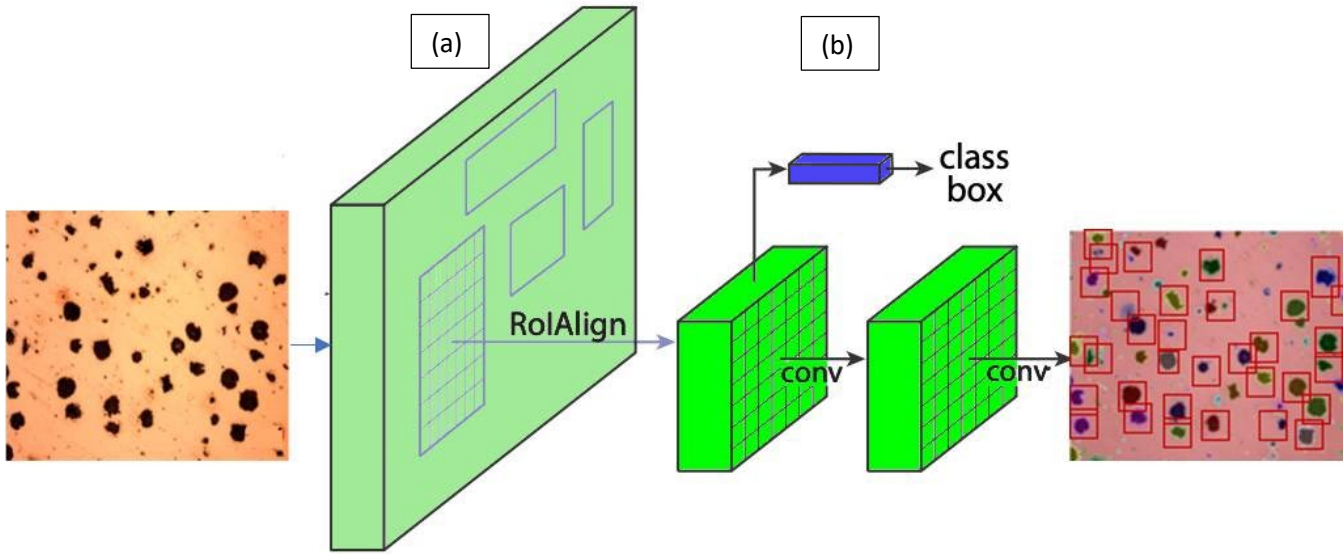

*Figure 11: General Mask R-CNN architecture for instance segmentation. (a) represents the Faster R-CNN architecture followed by (b) which represents two FCN layers. The output consists of object classification, regressor for the bounding box around the object, and mask classification [77].*

Li et al.[78] proposed a novel instance segmentation framework called prior Mask R-CNN by fusing prior knowledge for automatic precipitation detection. The proposed method is a two-stage instance segmentation network based on Mask R-CNN and the region of interest align (RolAlign) method for the measurement of precipitates in aluminum alloys. The dataset consists of six-series aluminum alloy with good formability, corrosion resistance, weldability, and low cost. In the experiment, direct chill cast Al–12.7Si–0.7Mg alloy was used with 30 metallographic slices with the size of 2048 × 2048 under different heat treatment conditions. From the final prediction, the method extracts specific measurements such as the length of precipitations. The method achieved the mean average precision of 0.475 on bounding box detection and 0.298 on mask segmentation tasks.

## 3.5 Image reconstruction and synthesis

Generative model is a powerful tool for microstructure representation and design. GAN techniques have been influential in material imaging, and it can be used to generate synthetic images of materials. Rühle et al. [30] proposed a workflow that can provide fully automated particle segmentation of agglomerated and non-spherical nanoparticles and find the particle size distribution from SEM images with minimal user interaction. Traditionally, human experts would be required to perform detection and find the objects of interest location, which is time-consuming. The proposed method used GAN to generate fake images and masks, in such a way that the model can be trained without a large set of labelled images. The workflow starts with the user providing SEM micrographs of $TiO_2$ particles and some representation of the outlines of the particle shapes. The rest of the steps are automatic. First, the workflow generates more particles using wGAN and combines them to create fake segmentation masks. Second, CycleGAN is trained on real SEM images with the fake masks from the previous step to create more fake images and their masks and to make them more realistic by removing any artifacts. Lastly, MultiResUNet is applied to extract the particle size distributions from the SEM images. The method gives results comparable to that of UNet trained with labelled images.

Yang et al. [8] proposed a GAN-based model for microstructural material design. The proposed method is used for an end-to-end solution for the low-dimensional latent space and non-linear microstructure characteristics, which learns the mapping between the latent design variables and microstructures. The proposed method was trained on 5000 synthetic microstructure images of size $128 \times 128$ created using Gaussian Random Field (GRF). The method was combined with a Bayesian optimization approach to design microstructures with optimal optical absorption performance. The results showed that the optical performance of the GAN-generated microstructures was 4.8% better than that of randomly sampled microstructures and was 17.2% better with Bayesian optimization. The authors concluded that the GAN method could generate arbitrarily-sized microstructures by changing the size of the latent variables and the discriminator can be used as a pretrained model for developing structure–property prediction models.

Chun et al. [79] proposed utilizing GAN to create ensembles of synthetic heterogeneous energetic material microstructures. The proposed GAN has two inputs: local stochasticity parameter $\rho$ which is the same as the latent variables in the standard GAN and the global morphology parameter $\lambda$ used to control the overall morphological characteristics of the generated image such as grain size, aspect ratio, and orientations. A synthetic microstructure generated by proposed GAN showed better qualitative and quantitative results compared to a synthetic image generated by transfer learning. Additionally, the microstructures produced by the proposed GAN were more similar to real microstructures images and statistically meaningful. The limitation of the method was that it could not generate crystals bigger than 125 pixels in size.

DNN have been used to estimate the phase volume fraction of the advanced high strength steel (AHSS). One drawback is that it requires a large amount of labelled data. To overcome of this challenge, Kim et al. [80] proposed an unsupervised DNN to estimate the phase volume fraction of multi-phase steel without the manual labelling. As shown in Figure 12, the proposed method uses an information maximizing generative adversarial network (InfoGAN) to learn the underlying probability distribution of each phase and to generate realistic sample points with class labels. It was followed by MLP classifier that used the generated data for the training. Finally, to estimate the phase volume fraction, the areas of the identically labelled samples were added. MLP is trained to learn a decision boundary and to predict labels. To validate the proposed method for AHSS alloys, they collected Electron Back-Scatter Diffraction (EBSD) microstructure images of six types of steels with different kinds of chemical compositions manufactured under different processing conditions. They used TIMS software to analyze and extract 19 features including grain size (diameter), grain average misorientation (GAM), and solidity from each segmented EBSD image. The training dataset consists of 116,811 samples without any labels. Since the dataset is numerical, 1D InfoGAN was implemented. The proposed method was used to estimate the true phase fraction of steel and the results showed a mean relative error of 4.53% at most, which means that the estimated phase fraction is close to true phase fraction.

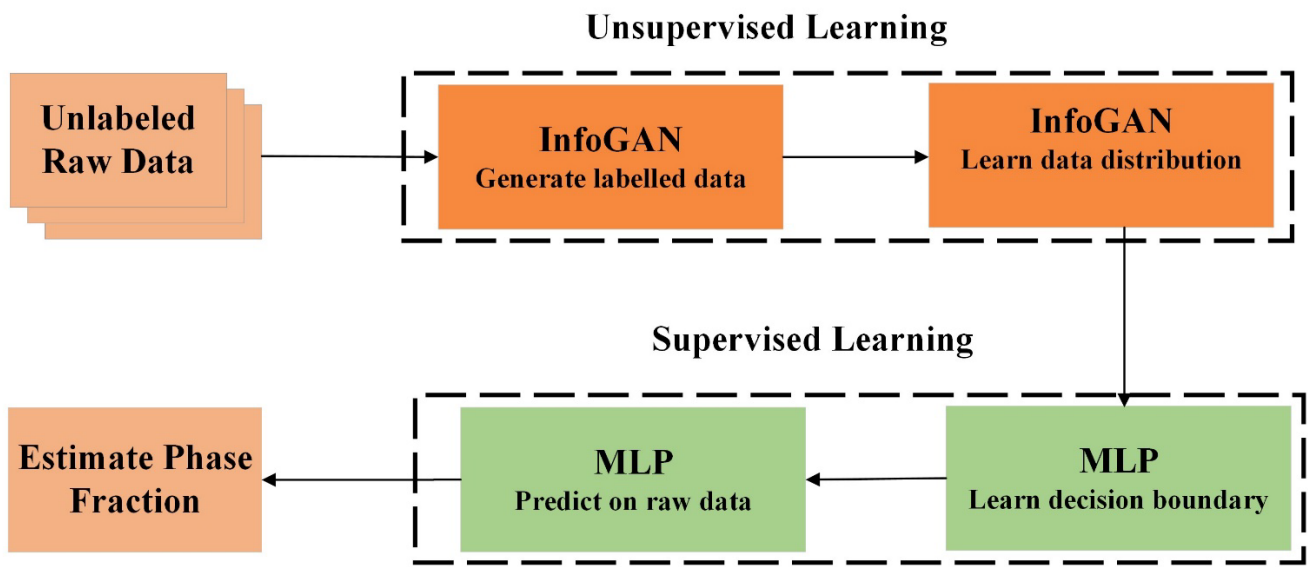

*Figure 12: The workflow of the DNN method. The workflow consists of two parts: the first part was the InfoGAN trained using unlabeled raw data, the second part was a MLP classifier which was trained to learn a decision boundary and predict labels on the raw data [80]*

Inverse material design requires navigating in chemical space through calculation and simulation [81]. VAE and GAN are particularly useful for inverse material design. For example, Long et al. developed an inverse design framework called *constrained crystals deep convolutional GAN* (CCDCGAN) that was used to design the crystal structure with low formation energy. They applied it to the binary Bi-Se system. They optimized DCGAN with some conditional variables to generate distinct crystal structures. They selected formation energy as the target property and performed density functional theory (DFT) calculation to generate sufficient training data [82]. Table 1 lists the latest research where CV methods were applied for various image analysis tasks.

Table 1. List of the literature in which CV are applied for microstructural analysis tasks, including image classification, semantic segmentation, object detection, instance segmentation, and image reconstruction and synthesis.

| Authors | Image analysis tasks and materials | Model architecture | Transfer Learning /Fine-tuning applied | Number of images (For training/ tuning) | Model Performance | Open access data |
|---|---|---|---|---|---|---|
| Azimi et al.[66] | Semantic Segmentation for classifying the microstructural constituents of low carbon steel | FCNN+ max-voting scheme | Yes | 21 images, 11 images for training and 10 images for testing | 93.94% classification accuracy | No |
| Furat et al [83] | Semantic segmentation for computing the tomography data of Al-Cu specimen. | 3D UNet 2D UNet | No | They have two types of datasets: 1- Simulated: 2769 samples for training, and 924 for testing  2-hand-labeled: 220 samples for training, and 74 for testing | Simulated: 82.1% accuracy.  hand-labeled: 73% accuracy | No |
| Saito et al. [84] | Segmentation and identification of the thickness of atomic layer flake for two-dimensional (2D) crystals. | UNet | No | 24 and 30 OM images of MoS2 and graphene, respectively | accuracy rate: 70% to 80% | Yes |
| Cabrera et al. [85] | Instance segmentation of yttrium silicate and silicon oxide | Mask R-CNN | No | 26 TEM images | MAP: 90% of accuracy | No |
| Liang [86] | Object detection for recognition and measuring the shape and size of ZnO | Mask R-CNN | Yes | 45 SEM images | area under the curve (AUC): 83.1% | No |
| Anderson et al. [87] | Object detection to detect helium bubbles in micrographs of | Faster R-CNN | Yes | 230 TEM images | Accuracy levels of 93% | Yes |
| | neutron-irradiated Inconel X-750 | | | | | |
| Shen et al. [88] | Object detection to extract defect size of different FeCrAl alloys | Faster R-CNN + watershed flood algorithm | Yes | 165 STEM images | F1 score of 78% | Yes |
| Roberts et al. [79] | Instance segmentation to segment and detect multiple defect types (Dislocation lines, precipitates and voids) in steel | DefectSegNet | No | Two large STEM images divided to 48 STEM images | IoU: 61.79 ± 2.13 % | No |
| Cohn et al. [89] | Instance segmentation for segmentation of each particle to generate detailed size and morphology information of a gas-atomized nickel superalloy powder | Mask R-CNN | Yes | SEM images. 1360 powder particle instances and 1029 satellite instances | precision and recall of 0.938 and 0.799 | sample |
| DeCost et al. [65] | semantic segmentation Two tasks: 1- semantic segmentation of steel micrographs into four regions (grain boundary carbide, spheroidized particle matrix, particle-free grain boundary denuded zone, and Widmanstätten cementite) 2- segmenting cementite particles within the spheroidized particle matrix | PixelNet CNN | Yes | 20 images for training and 4 images for testing for from UHCS database (2D SEM micrograph of ultrahigh carbon steel) | Precision: 92.6 Recall: 92.6 IoU: 75.4 | Yes |
| Siao et al. [90] | semantic segmentation to detect graphene from microscopic images | Automatic Graphene Detection Method with Color Correction (MLA-GDCC) (Modified U-Net + SVM) | No | 246 OM images for training and 57 OM images for testing | pixel-level accuracy: 1- monolayer: 99.27 2- bilayer:98.92 | No |

| Reference | Task | Model | Data Augmentation | Dataset | Performance | Open Source |
|---|---|---|---|---|---|---|
| Durmaz et al.[91] | Semantic segmentation to predict lath-bainite segmentation | UNet | Yes | 30–50 2D micrographs of different imaging modalities and electron backscatter diffraction-informed annotations (SEM or LOM) | IoU: 71.6 ± 1.7 | Yes |
| Mianroodi et al.[92] | Semantic segmentation for local stress calculations in inhomogeneous non-linear materials | modified UNet | No | ——— | MAPE for elasto-plastic materials: 6.4% | No |
| Li et al. [93] | Object detection for detecting multiple defect types using images of varying contrast, brightness, and magnification | cascade object detector + CNN (15 layers) + image processing concepts | No | 298 STEM micrographic images | recall: 0.858 precision: 0.865 | No |
| Maksov et al.[94, p. 2] | Classification. DL framework for dynamic STEM imaging to find the lattice defects. It was applied for mapping solid state reactions and transformations in layered WS2, and classification into different categories using unsupervised clustering methods. | adapting Gaussian mixture model (GMM) on the CNN (encoder-decoder) | No | ——— | accuracy on the test set ~99% | No |
| Pokuri et al. [95] | Classification to prediction short-circuit current from morphology maps of organic photovoltaic films. | Deep Learning for Structure Property interrogation (DLSP), VGG16 and ResNet-50 | Yes | 65,000 morphologies | accuracy: VGG16: 96.61% ResNet-50: 96.45% DLSP: 95.80% and F1-score 97.28% | Yes |
| Kim et al.[96] | Semantic segmentation to | Unsupervised CNN+ | Yes | The dataset contains 7 | ——— | No |
| | segment a low-carbon steel microstructure (Ferrite, pearlite, and martensite) | superpixel algorithm | | microscopy images. | | |
| Jung et al.[97] | Generative model to generate a super-resolved microstructure image | super-resolution (SR) residual network (SRResNet) | No | 2D and 3D SEM and EBSD images. 12,000 pairs of low-resolution and high-resolution images | ——— | No |
| Horwath et al.[98] | Segmentation of high-resolution ETEM images | CNN (encoder-decoder, or 'hourglass" type CNN architecture) + UNet | No | 2400 full ETEM images | Mean F1 score for UNet ~= 91% | No |
| Lin et al. [99] | Instance segmentation to segment vanadium pentoxide (V2O5) nanowires | Mask R-CNN | No | 1000 synthetic 3D SEM, X-ray ptychography, and STXM images | X-ray ptychography accuracy: 86.6% STXM accuracy: 75.6 SEM accuracy: 61.7 | Yes |
| Tong et al.[100] | Semantic segmentation to segment carbon fibers (CFs) and CF clusters from SEM images for quantitative CF distribution characterization | FCN | No | 560 CFRC 2D SEM images. | F1-scores 0.94 precision:0.96 recall:0.92 | No |
| Shin et al. [101] | Object detection to identify and classify the graphene and hBN with different sizes, shapes and thicknesses form silicon oxide wafer | graphene trained deep neural network (GT-DNN) | Yes | 1134 2D OM images, graphene (1006 images) hBN (128 images) | Average precision (AP): 80.23% | No |
| Han et al.[102] | Semantic segmentation to demonstrate the material and thickness | An encoder-decoder NN called 2D material | Yes | 817 OM images containing exfoliated flakes of 13 materials | IoU: 58.78% | No |

| | identification of 13 different 2D materials, including graphene/graphite, hBN, 2H-MoS, 2H-WS$_2$, 2h-WSe$_2$, 2HMoTe$_2$, 2H-TaS$_2$, 2H-NbSe$_2$, 1T-HfSe$_2$, black phosphorous (BP), CrI$_3$, RuCl$_3$ and crystal ZrTe$_5$ | optical identification neural network (2DMOINet) | | and 100 background-only images | | |
|---|---|---|---|---|---|---|
| Ushiba et al. [103] | Semantic segmentation to classify graphene | UNet | No | 808 OM images | F1-scores exceeding 80% | No |
| Yildirim and Cole [104] | Instance segmentation to segment particle instances such as silica, Au nanorods, Pt$_3$Co, Pd, and Fe$_3$O$_4$ | Bayesian deep-learning model | No | 465 EM images | mIoU: 0.844 | Yes |
| Groschner et al. [105] | Semantic segmentation to detect the stacking faults in CdSe and Au | UNet + RF | No | 2D HRTEM images. 46 micrographs of CdSe, and 13 micrographs of Au | UNet DICE:0.80 precision: 0.82 recal:0.78, RF Accurcy: 86% | Yes |
| Torbati-Sarraf et al. [106] | Semantic segmentation to detect particle nodules of Al-Cu-Fe, Cu-Mg-Zn | U-Net, UNet++, PSPNet, DeepLab v3 | No | 3D Transmission x-ray Microscopy (TXM) | mIOU U-Net: 92.2% UNet++: 95.1% PSPNet: 88.2% DeepLabv3+: 89.1 | No |

## 4. Challenges and Future Directions of CV for Microstructural Analysis:

There are several challenges in using CV for microstructural analysis, which include data preparation and deciding the specific image analysis task needed for a given problem. This section lists the challenges and several ways to overcome these challenges.

## 4.1. Preparing Dataset for Training

*a) Noisy raw data*: As shown in figure 13 (a), the microstructural images often contain noises such as salt-and-pepper noise (sparsely occurring white and black pixels), Gaussian noise, blurring problems, and low contrast. These issues can be considered one of the major complicating factors for CV algorithms. Different image-processing techniques have been used to resolve these issues. For examples, anisotropic diffusion was used to reduce image noise without removing significant parts of the image content compared to traditional filtering algorithms that might cause blurring of relevant details [50], [107]–[109].

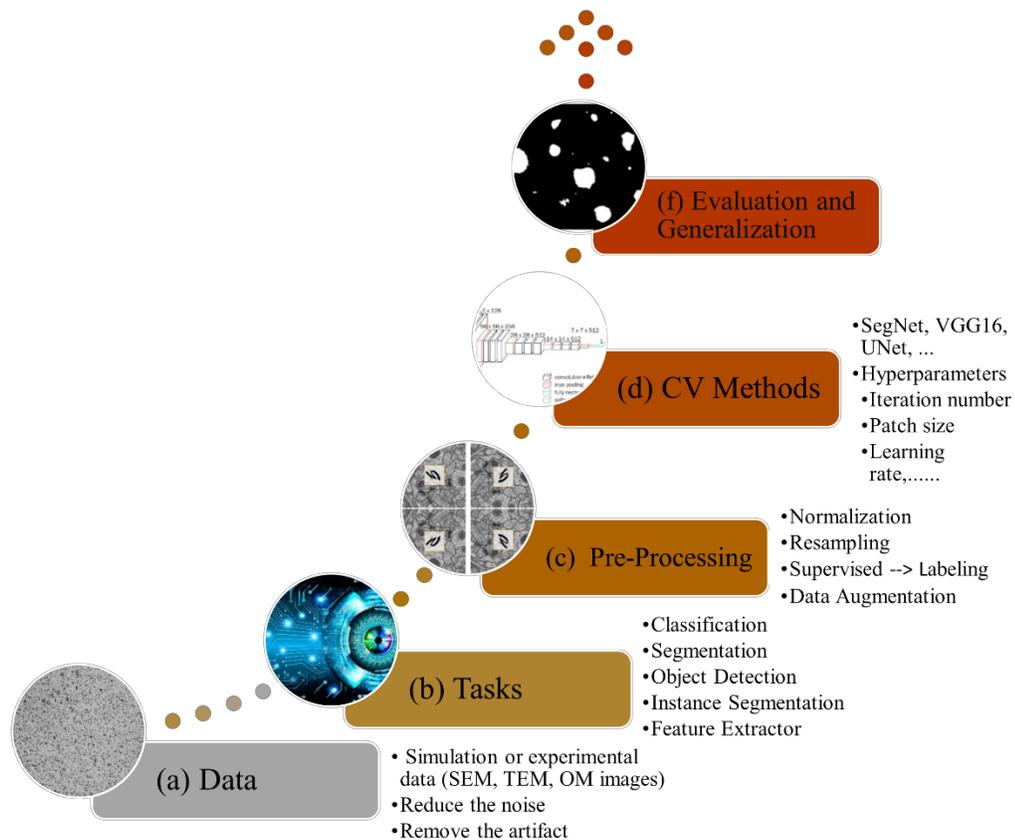

*Figure 13: The general workflow of computer vision tasks.*

*b) Unlabeled dataset*: Extracting different kinds of information from microstructure images depends on image analysis tasks as shown in figure 13 (b) [5]. In supervised learning, the training dataset contains images and labels to train the model by examples. For instance, in semantic segmentation, labels provide the ground-truth of the semantic classes of the image pixels. The labels can be generated manually by material experts who annotate images using software such as ImageJ, GIMP (GNU Image Manipulation Program), label Studio, and QuPath. The labels can also be generated using ML techniques such as K-means [64]. Potential problems with image labelling include boundary distortion and interpolated edges. These problems may be mitigated by coding or postprocessing to reduce interpolating between regions or classes.

*c) Small dataset*: Another challenge in CV is that it usually needs large dataset to achieve good performance. However, the datasets in materials research are usually small and sometimes more diverse than data in other fields. To address this problem, there are three possible solutions:

1- Data augmentation: It provides more samples in the training set, which reduces the risk of overfitting and improves the generalization of CV methods [48] [90]. Data augmentation techniques include image transformations such as random cropping, flipping the image horizontally or vertically, resizing, and translation. Albumentations library [110], Tensorflow-Keras, and CV methods such as Generative Models [111] can be used to provide more samples to the training set as shown in Figure 13 (c). For example, Alrfou et al.[64] used Albumentations library to generate more than 500 samples for the training step while the original dataset contains only 10 images for training and 3 images for testing. Stan et al. [112] also used image augmentation to train SegNet to predict the solidification dendrite.

2- Transfer-learning and fine-tuning: Transfer learning includes training the DL model (the source domain) with a large amount of data and taking the knowledge gained from source domain by modifying the last part of the network for a new task while freezing the feature extraction part. Fine-tuning is similar to transfer learning except that the fine-tuning updates the parameters on the feature extraction part as well. Both techniques can be used to train the model (target domain) with a relatively small dataset [113]. CNN models such as AlexNet, GoogleNet, ResNet, and VGG have been trained on large datasets such as ImageNet [17] and MS-COCO [73] for image classification tasks. The wights and parameters can be transferred or fine-tuned to a new model, which reduces the amount of time for training the new model from scratch with smaller datasets. Transfer learning has also been widely used for analyzing materials microstructure images. For example, Modarres et al. [114] applied CNN with transfer learning (using Inception-V3 pre-trained on the ImageNet 2012) for SEM images of nanostructures. They used around 20,000 SEM images that were manually classified into 10 categories (Porous_Sponge, Patterned_surface, Particles, Films_Coated_Surface, Powder, Tips, Nanowires, Biological, MEMS_devices_and_electrodes, and Fibers). CNN-based model was then used for automatic categorization and labelling of images.

Li et al. [115] proposed a transfer learning-based approach using pre-trained VGG19 for reconstructing statistically equivalent microstructures from a wide range of material systems. The transferred deep convolutional network is pruned to reduce the computational cost by dropping the higher-level convolutional layers. The knowledge learned in model pruning can be transferred to developing structure-property predictions.

3- Simulation: Collecting experimental data is time consuming and requires human expertise. Simulation has lower cost and can be used to provide more data. Table 3 shows a list of available datasets generated by experiments or simulations that can be applied to CV tasks. For example, CrystalKit is a program used to build crystalline defects of various kinds from point defects to grain boundaries and precipitates. Landyne 2 published in 2016 is used as a practical tool in research and as teaching aid on microscopy and crystallography. It can be used for generating electron diffraction patterns. However, in some cases, the simulated data may carry less information than the experimental data. To address this issue, Ma et al. [116] used Monte Carlo Potts model to represent the polycrystalline microstructure of materials images and labels. The method extracts useful features from the real images and modify the simulated images by applying GAN to generate synthetic images and make them more relevant to the segmentation task.

*Table 1: List of some experimental/simulation Datasets*

| Experimental/ Simulation Dataset | Description |
| --- | --- |
| Atomagined [117], [118] | A simulated atomic-resolution high angle annular dark field (HAADF) imaging dataset containing unique ICSD structure prototypes. |
| DeepDamage [119] | Deep learning to quantify and classify damage in dual phase steel |
| UHCSDB [120] | Ultra-High Carbon Steel Micrograph Database |
| UHCS [65] | Ultrahigh carbon steel (UHCS) micrographs |
| Warwick electron microscopy dataset [121] | TEM and STEM Images/Crops datasets |
| Powder bed anomaly [122] | The dataset was collected and annotated using the internally developed Peregrine software tool. The dataset consists of three different types of layer-wise powder bed images |

| | including: laser powder bed fusion, electron beam powder bed fusion, and binder jetting. |

*d) Imbalanced dataset*: Another challenge is the imbalanced dataset in materials microstructures, where there are far more negative samples than positive ones. Alrfou et al. [64] used an imbalanced dataset, where most of the pixels belonged to the background class (Iron) as opposed to the particle (graphite) pixels. They used focal-loss function [123], which generalizes the binary cross-entropy loss, to improve the imbalance dataset by assigning more weights to hard misclassified examples. Zhou et al. [124] proposed novel defect detection model of material microscope image called Material Defect Detection Network (MDD-Net), where focal-loss function is also applied to address the imbalance between foreground and background.

## 4.2. Explainability

Deep learning models are often considered as black boxes, where the processing mechanism of the internal structure does not explain why a given input produces a specific output. In CNN models, each convolutional block uses convolution function with multiple kernels or filters. These filters have relatively small number of columns and rows and they convolve over the input to extract high-level local features in the form of feature maps [125]. Visualizing feature maps in CNN can provide some insights about the prediction process. Figure 14 shows the features maps for the second and fifteenth layers in a UNet model for the segmented graphite nodules in ductile iron microstructure [64]. In the second layer, different filters were applied to extract the feature maps representing edges, textures, and blurring, which is human readable. However, going deeper in the network, the filters concentrate more and learn more abstract features. As shown in Figure 14, in the last down-sampling layer (layer 15), feature maps are very abstract and cannot be interpreted by human.

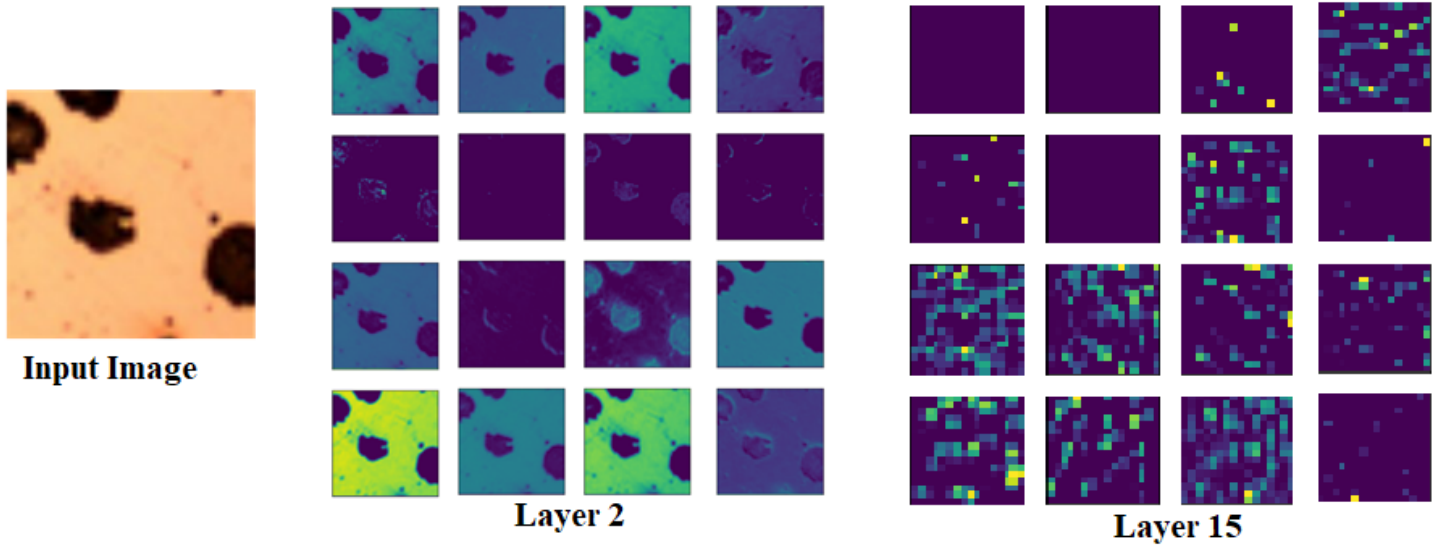

*Figure 14: Visualization of a 2D image of graphite in ductile iron, and the resulting 16 feature maps in the second convolutional layer, and 16 feature maps in the last convolutional layer in the encoder of the UNet architecture.*

To address the black-box nature of the CNN models, a number of studies have been conducted to develop interpretation techniques and tools in order to explain and visualize the decision-making process [126], [127]. Computing the heatmap is the most popular approach for visualizing and understanding CV tasks. The heat map is the class activation map, highlighting the importance of the image region to provide more details of why the model predicts a specific output. Zhou et al. [128] proposed a Class Activation Mapping (CAM) method for explaining CNN image prediction classifiers. This technique generates a CAM using the global average pooling (GAP) and adds additional layers on the top of GAP to learn classes specific in CNN. CAM obtains the weighted average of the last convolutional features using the fully connected layer weights for each class of the network. For each class, the CAM indicates which parts of the image are effective in classification. The drawback of CAM is that it requires GAP, which cannot apply to every CNNs. Selvaraju et al. [129] proposed gradient-weighted CAM (Grad-CAM) to solve the limitation of CAM on CNN architecture and introduced the gradient information to CAM. The proposed method generalizes well for most CNNs. Zeiler & Fergus [130] proposed deconvolutional network (deconvnet) that performs an inverse convolution model. The method is used to highlight the part of the image that is responsible for the activation of each neuron and show what input pattern originally causes a given activation in the feature maps. In another research, an integrated gradients optimization saliency (I-GOS) and I-GOS++ were developed to avoid local optima in heatmap generation, which advanced single image explanations by improving the performance across all resolutions [131], [132].

Visualization methods have been used to interpret the CV methods applied to classify material microstructure images. Xing et al. [133] used vanilla CNNs and four state-of-the-art CNN models AlexNet, ResNet, SqueezeNet, and InceptionV3 to classify solidified melt pool images captured in a pulsed selective laser melting (P-SLM) process. They employed the images captured at five input laser energy levels to train and test CNN models. To avoid overfitting, data augmentation and dropout layers were integrated into the model. The experiments were performed using Q235A steel powder with an average diameter of 27μm. The results show that ResNet has improved the classification accuracy of melt pools to 96.6%, which is resulted from the ability to utilize complex image features from the melt pool region. To ensure the classification results are reasonable, three visualization methods were used to add interpretation of the AlexNet model: CAM, deconvnet, and Grad-CAM. The visualization methods show that the melt pool area has higher weights than other areas such as the background and noise.

**4.3. Transformers for microstructural analysis:**

Transformer based models can be applied to various computer vision tasks such as image classification, image segmentation, object detection, and image synthesis domains. The existing transformer-based models are mainly trained with 2D natural images such as ImageNet and other large-scale datasets for pre-training and fine-tuning. It is shown that transformer models have achieved impressive results over benchmark datasets including medical datasets compared to CNN-based models. This is due to the fact that the original transformer model ViT [38] was able to capture long-range dependencies between pixels and global information. However, ViT still has poor performance in the extraction of local information. Several transformer models were introduced to help capturing local information. In medical segmentation, the transformer-based models are pre-trained on ImageNet for the fine-tuning task [69]. These approaches have been sub-optimal due to the large domain gap between natural and medical images. Recent GAN transformer-based approaches can effectively generate more realistic images than GAN CNN-based models and avoid the limitation of GAN CNN models such as instability.

A few research projects have been conducted for image segmentation of materials data using Swin-UNet++ [68]. Since the method can capture the global and long-range dependency, it is effective in detecting and identifying the boundaries and interior regions of dimples micro-features with complex morphology from the input of 2.25Cr1Mo0.25V steel image. To the best of our knowledge, no work has been done using transformer-based models for object detection, image classification, and image synthesis. We believe that the transformer models in both pure transformer and hybrid architecture can be adopted for material imaging and they may demonstrate better performance compared to CNN-based models. This is due to their ability to capture long-range dependencies and global information between pixels, which might capture complex structures and irregular shapes and improve the prediction results. We believe that the training datasets will play a critical role in exploring the limits of transformers for CV in material science. Since transformer-based models are pre-trained on natural images, the features detected do not translate well to material microstructural analysis due to the significant semantic gap between natural images and microstructural images. Creating a large labelled dataset for microstructural images and pre-training the transformer models using the dataset will improve the model performance.

## Conclusion

This paper highlighted the latest computer vision (CV) methods for the microstructural analysis of material. We briefly reviewed the concepts behind the success of convolutional neural network models and discussed the tradeoffs of various models and their impact on material sciences. We characterized the main image analysis tasks i.e., classification, semantic segmentation, object detection, instance segmentation, feature extraction, and image reconstruction that can be performed by variations of CNN algorithms. We underscored the key strengths and weakness of the existing methods. Moreover, some recent applications in designing new materials were highlighted, including predicting complex stress and strain fields in composite materials, identifying, and classifying particles with different sizes, shapes, and thicknesses, and inverse designs.

Despite the success of CNN for materials characterization, the approach is still in its early stage and has great potential for the discovery, design, and development of new materials. The availability of large training datasets is critical to the success of applying CV methods to material science. Other challenges pertain to the presence of noisy raw data, unlabeled data, and imbalanced dataset. Some solutions to resolve these issues have been suggested.

Finally, we identified two main venues for future research about the application of CV methods applied to materials science. First, to develop new techniques to enhance explainability of the learning process of CNNs to ensure the output of the algorithms are interpretable by the underlying physics. Second, we covered the concepts of transformer models and how to adapt these techniques to material science. In several fields, including health science, it has been observed that there is a growing interest in transformer-based models. This is due to the fact that the transformer-based model identifies long-range pixel relationships, leading to a global understanding of images, while CNN models operate locally. We discussed that through the synergy of the self-attention mechanism with CNN models, local and global features can be captured, which can outperform the traditional CNN models.

**Conflict of interest:** All authors declare no conflict of interest.